\documentclass[10pt]{iopart}

\usepackage{enumerate}
\usepackage{array}
\usepackage{amssymb}

\makeatletter                                                
\newcount\c@MaxMatrixCols                                       %
\def\matrix{\hskip -\arraycolsep \array{*\c@MaxMatrixCols c}}   
\let\@xp=\expandafter                                           %
\makeatother                                                 

\eqnobysec

\def\abs#1{\left \vert #1 \right \vert}
\def\totdiff#1#2{\frac{\rmd #1}{\rmd #2}}
\def\parcdiff#1#2{\frac{\partial #1}{\partial #2}}
\def\sparcdiff#1#2{\frac{\partial^2 #1}{\partial #2^2}}

\begin{document}

\title{Painlev{\'{e}} test of coupled Gross-Pitaevskii equations}
\author{D{\'{a}}niel Schumayer and Barnab{\'{a}}s Apagyi}
\ead{\mailto{apagyi@phy.bme.hu}, \mailto{schumayer@phy.bme.hu}}
\address{Institute of Physics, Budapest University of Technology
and Economics, \newline
H-1111, Budafoki {\'{u}}t 8, Hungary}

\begin{abstract}
Painlev{\'{e}} test of the coupled Gross-Pitaevskii equations has
been carried out with the result that the coupled equations pass
the P-test only if a special relation containing system parameters
(masses, scattering lengths) is satisfied. Computer algebra is
applied to evaluate $j=4$ compatibility condition for admissible
external potentials. Appearance of an arbitrary real potential
embedded in the external potentials is shown to be the consequence
of the coupling. Connection with recent experiments related to
stability of two-component Bose-Einstein condensates of Rb atoms
is discussed.
\end{abstract}

%

\section{Introduction}

Recently  there has been a growing interest in the
Gross-Pitaevskii (GP) equations \cite{Gross1961,Pitaevskii1961}
describing two-component Bose-Einstein condensates (BEC) in
external trap potentials [3-22]. In the absence of the confining
potential, the GP equations reduce to the coupled non-linear
Schr{\"{o}}dinger (NLS) equations which play an important role in
optics \cite{Hasegawa1990}. Coupled GP equations are also used to
describe Josephson-type oscillations between two coupled
BEC~[9-11], 
spin-mixing dynamics of spinor BEC~[12-15], 
or to explore such interesting field of matter waves as possible
atomic soliton lasers \cite{Ballagh1997,Ruprecht1995,Heinzen2000}.

An efficient tool of the analysis of the non-linear partial
differential equations is the Painlev{\'{e}} (P) method
\cite{Weiss1983,Steeb1988} which serves to explore the singularity
structure of the underlying equations, and establish integrability
conditions \cite{Ablowitz1991}. The P-analysis of the single NLS
equation has been performed by Steeb \etal \cite{Steeb1984}, and
the damped NLS (or the GP) equation has been investigated by
Clarkson \cite{Clarkson1988}. A fairly large class of coupled NLS
equations including third order dispersions have been analyzed by
Radhakrishnan \etal \cite{Radhakrishnan1995}. Recently the
symmetrically coupled higher-order NLS equations have been tested
by using the P-method \cite{Sakovich2000}.

Because of the experimental developments in forming two-component
BEC \cite{Matthews1999} and the possibility to confine BEC in a
linear shape \cite{Fortagh1998}, we shall perform the P-test of
the coupled one-dimensional GP equations in order to establish
certain necessary conditions of integrability. (The term
integrability is used here in the general sense
\cite{Steeb1988,Ablowitz1991} involving P-property and soliton
formation.) The results obtained for the trap potentials are
similar to those found by Clarkson \cite{Clarkson1988} in the case
of the damped NLS equations: the trap potential should be linear
and/or quadratic in the coordinate variable $x$. In the quadratic
case a source term depending only on time $t$ should also be
present in the external potential $V(x,t)$.

A novel feature of our analysis is the possibility of the
appearance of an arbitrary common potential term $\tilde{V}(x,t)$
within the confining potentials $V_1(x,t)$ and $V_2(x,t)$. Its
presence may prove useful for fine tuning experiments with two
component BEC. We consider the system of coupled GP equations in
its most general form containing different masses, external
potentials, and mutual coupling strengths. As a result we shall
derive compatibility conditions, the fulfillment of which depends
on the parameters characterizing the GP equations. We show that in
a particular experiment \cite{Hall1998_1} employing two hyperfine
states of Rb atoms as components of BEC, the vortex stability
corresponds to just the parameter ratios satisfying our general
formula derived in this paper.

The organization of this paper is as follows. In section 2 the
P-analysis of two coupled GP equations will be carried out
including the determination of the leading orders, the recursion
relations, the resonances, and the compatibility conditions. The
consequences of the compatibility relations for the potentials are
discussed in section 3 where also other consistency requirements
are studied.  In section 4 we make comparisons with earlier
results and investigate compatibilities with existing experimental
and numerical findings related with  two-component BEC. Section 5
is devoted to a short summary.

\section{Painlev{\'{e}} test \label{sec:Painleve_test}}

Let us consider the following $(1+1)$ dimensional inhomogeneous
NLS equations for the wave functions $\psi_{1}$, $\psi_{2}$ with
the external potentials $U_1(x,t)$, $U_2(x,t)$ \numparts
\begin{eqnarray} \label{eq:csatolt_GP_fizikai_jelolessel}
  \fl  \rmi \hbar \parcdiff{}{t} \psi_{1}(x,t)\!\! =
       \!\!\left ( - \frac{\hbar^{2}}{2m_{1}} \nabla^{2} +
       U_{1}(x,t) + U_{11} \abs{\psi_{1}(x,t)}^{2} +
       U_{12} \abs{\psi_{2}(x,t)}^{2} \right )
       \psi_{1}(x,t) + U_{10}, \nonumber \\
       \\
  \fl  \rmi \hbar \parcdiff{}{t} \psi_{2}(x,t)\!\! =
       \!\!\left ( - \frac{\hbar^{2}}{2m_{2}} \nabla^{2} +
       U_{2}(x,t) + U_{21} \abs{\psi_{1}(x,t)}^{2} +
       U_{22} \abs{\psi_{2}(x,t)}^{2} \right )
       \psi_{2}(x,t) + U_{20} \nonumber \\
\end{eqnarray}
\endnumparts
which, in the absence of the inhomogeneities $U_{10}$ and
$U_{20}$, are commonly called  the coupled Gross-Pitaevskii
equations \cite{Gross1961,Pitaevskii1961}.

Here $m_i$ denotes the mass of the atomic species $i$ $(i=1,2)$ of
the two-component BEC gas and $U_{ij}$ is related with the
interactions between the atoms $i$ and $j$ $(i,j=1,2)$ via the
relation $U_{ij}=2\pi\hbar^2a_{ij} N_{j}/A\,\mu_{ij}$ where $N_j$
means the number of atoms in the $j$th component of the BEC,
$a_{ij}$ is the scattering length characterizing the interaction
between atoms $i$ and $j$, $A$ represents a general cross
sectional area confining species $i$ and $j$, and $\mu_{ij}=m_{i}
m_{j}/(m_{i} + m_{j})$ is the reduced mass.

By introducing the new parameters
\numparts
\begin{equation}\label{eq:parameters}
    \lambda   = \frac{\hbar}{2m_{1}},   \quad
    \vartheta = \frac{\hbar}{2m_{2}},   \qquad
    T_{ij}    = \frac{1}{\hbar} U_{ij}, \quad (i,j = 1,2),
\end{equation}
and notations
\begin{equation}
    u = \psi_{1} \, , \quad
    w = \psi_{2} \, , \quad
    V_{i} = \frac{1}{\hbar}\, U_{i} \, , \quad
    V_{i0} = \frac{1}{\hbar}\, U_{i0} \, , \quad (i=1,2)
\end{equation}
\endnumparts
we write the GP equations into the standard form of the P-analysis
\numparts
\begin{eqnarray} \label{eq:GP_mat}
    \rmi u_{t} + \lambda u_{xx} - T_{11} \abs{u}^{2} u -
    T_{12} \abs{w}^{2} u &=& V_{1} u + V_{10}, \\
    \rmi w_{t} + \vartheta w_{xx} - T_{21} \abs{u}^{2}\! w -
    T_{22} \abs{w}^{2} w \!&=& V_{2} w + V_{20}
\end{eqnarray}
\endnumparts
where $T_{ij}$ and $\lambda$, $\vartheta$ stand, as defined by
equation \eref{eq:parameters}, for the interaction and mass
parameters, respectively.

In order to apply the P-analysis, we first complexify all
variables to obtain equations (\ref{eq:GP_mat}\textit{-b}) in the
form ($v=u^{\ast}$, $z=w^{\ast}$):
\numparts
\begin{eqnarray} \label{eq:GP_komp_a}
    \rmi u_{t} + \lambda u_{xx} - T_{11} u^{2} v -
    T_{12} wzu &=& V_{1} u + V_{10}, \\
    \label{eq:GP_komp_b}
   -\rmi v_{t} + \lambda v_{xx} - T_{11} u v^{2} -
    T_{12} wzv &=& V_{1}^{\ast} v + V_{10}^{\ast}, \\
    \rmi w_{t} + \vartheta w_{xx} - T_{21} u v w -
    T_{22} w^{2} z &=& V_{2} w + V_{20}, \\
   -\rmi z_{t} + \vartheta z_{xx} - T_{21} u v z -
    T_{22} w z^{2} &=& V_{2}^{\ast} z + V_{20}^{\ast}
\end{eqnarray}
\endnumparts
where the functions $u$, $v$, $w$, $z$ are treated as independent
complex functions of the complex variables $x$ and $t$, and
$V_1^{\ast}(x,t)$, $V_{10}^{\ast}(x,t)$, $V_2^{\ast}(x,t)$,
$V_{20}^{\ast}(x,t)$ are formal complex conjugates of $V_1(x,t)$,
$V_{10}(x,t)$, $V_2(x,t)$, $V_{20}(x,t)$, respectively.

The next step is to seek the solutions of (\ref{eq:GP_komp_a}\textit{-d}) in
the form
\numparts
\begin{eqnarray} \label{eq:uvwz_sorf}
   \fl u(x,t) = \phi^{p} (x,t) \sum_{j=0}^{\infty}{u_{j}(t)
                \phi^{j}(x,t)}, &\qquad&
       v(x,t) = \phi^{q} (x,t) \sum_{j=0}^{\infty}{v_{j}(t)
                \phi^{j}(x,t)}, \\
   \fl w(x,t) = \phi^{r} (x,t) \sum_{j=0}^{\infty}{w_{j}(t)
                \phi^{j}(x,t)}, &\qquad&
       z(x,t) = \phi^{s} (x,t) \sum_{j=0}^{\infty}{z_{j}(t)
                \phi^{j}(x,t)},
\end{eqnarray}
\endnumparts
with the Kruskal ansatz
\begin{equation} \label{eq:Kruskal_ansatz}
   \phi(x,t)=x-\xi(t),
\end{equation}
and $\xi(t)$, $u_{j}(t)$, $v_{j}(t)$, $w_{j}(t)$, $z_j(t)$,
$j=0,1,2,\dots$ being analytic functions of $t$ in the
neighbourhood of a noncharacteristic movable singularity manifold
defined by $\phi=0.$ Similarly, the external potentials $V_i$
confining specimen $i$ is also expanded about the singularity
manifold $\phi=0$ as follows ($i=1,2$)
\begin{equation} \label{eq:V_sorf}
    \hspace*{-1cm}
    V_{i} (x,t) = \sum_{j=0}^{\infty}{V_{i,j}(t)
                  \phi^{j}(x,t)},\qquad
    V_{i,j} (t) = \frac{1}{j!} \left (
                  \frac{\partial^{j} V_{i}(x,t)}{\partial x^{j}}
                  \right )_{x= \xi(t)} \, .
\end{equation}

Substituting expansions (\ref{eq:uvwz_sorf}\textit{-b}) and
(\ref{eq:V_sorf}) into equations (\ref{eq:GP_komp_a}\textit{-d})
and equating like powers of $\phi$ we obtain:
\begin{enumerate}[i.)]
   \item{equations for determining the leading orders
         $p$, $q$, $r$, $s$;}
   \item{recursion relations for deriving the functions
         $u_{j}$, $v_{j}$, $w_{j}$, $z_{j}$.}
\end{enumerate}
In order that equations (\ref{eq:GP_komp_a}\textit{-d}) pass the
Painlev{\'{e}} test it is required that the numbers $p$, $q$, $r$,
$s$ be non-positive integers. Moreover, the recursion relations
should be consistent in all order of $j$ including the resonances.

\subsection{Determination of the leading orders}

To determine the leading orders $p$, $q$, $r$, $s$ appearing in
the expansions (\ref{eq:uvwz_sorf}\textit{-b}), it is sufficient
to consider the expansion upto the zeroth order, $j=0$. By
substituting this truncated version of expansions
(\ref{eq:uvwz_sorf}\textit{-b}) into (\ref{eq:GP_komp_a}) we
obtain \numparts
\begin{eqnarray} \label{eq:(2.7a)}
    \fl \rmi u_{0,t} \phi^{p} + \rmi u_{0}p \phi^{p-1} \phi_{t}
        + \lambda u_{0} p(p-1) \phi^{p-2} - T_{11} u_{0}^{2}
    v_{0} \phi^{2p+q} - T_{12} u_{0} w_{0} z_{0} \phi^{p+r+s}
    \nonumber \\
    \hspace*{6.7cm} = V_{1} u_{0} \phi^{p} + V_{10}.
\end{eqnarray}
Three completely similar expressions arise from the substitution
of the truncated version of expansions
(\ref{eq:uvwz_sorf}\textit{-b}) into the remaining three equations
(\ref{eq:GP_komp_b}\textit{-d}):
\begin{eqnarray}\label{eq:(2.7b)}
    \fl - \rmi v_{0,t} \phi^{q} - \rmi v_{0}q \phi^{q-1} \phi_{t}
        + \lambda v_{0} q(q-1) \phi^{q-2} - T_{11} v_{0}^{2} u_{0}
          \phi^{2q+p} - T_{12} v_{0} w_{0} z_{0} \phi^{q+r+s}
          \nonumber \\
        \hspace*{6.7cm} = V_{1}^{\ast} v_{0} \phi^{q} + V_{10}^{\ast} \, , \\
    \fl \rmi w_{0,t} \phi^{r} + \rmi w_{0}r \phi^{r-1} \phi_{t} +
        \vartheta w_{0} r(r-1) \phi^{r-2} - T_{21} u_{0} v_{0}w_0
        \phi^{p+q+r} - T_{22} w_{0}^2 z_{0} \phi^{2r+s}
        \nonumber \\
        \hspace*{6.7cm} = V_{2} w_{0} \phi^{r} + V_{20} \, , \\
    \fl - \rmi z_{0,t} \phi^{s} - \rmi z_{0}s \phi^{s-1} \phi_{t} +
          \vartheta z_{0} s(s-1) \phi^{s-2} - T_{21} u_{0} v_{0}z_0
          \phi^{p+q+s} - T_{22} w_{0}z_0^2 \phi^{r+2s}
          \nonumber \\
        \hspace*{6.7cm} = V_{2}^{\ast} z_{0} \phi^{s} + V_{20}^{\ast} \, .
\end{eqnarray}
\endnumparts

By demanding the leading order terms of equations
(\ref{eq:(2.7a)}\textit{-d}) to vanish one obtains the following
equations \numparts
\begin{eqnarray} \label{eq:(2.8)}
    \lambda p(p-1) &=& T_{11} u_{0}v_{0} + T_{12} w_{0}z_{0},    \\
    \lambda q(q-1) &=& T_{11} u_{0}v_{0} + T_{12} w_{0}z_{0},    \\
    \vartheta  r(r-1) &=& T_{21} u_{0}v_{0} + T_{22} w_{0}z_{0}, \\
    \vartheta  s(s-1) &=& T_{21} u_{0}v_{0} + T_{22} w_{0}z_{0}
\end{eqnarray}
\endnumparts \vspace{-4mm}
and
\numparts
\begin{eqnarray}
    p+q = -2, \\
    r+s = -2
\end{eqnarray}
\endnumparts
from which the leading orders can uniquely be determined to be
\begin{equation}
   p = q = r = s = -1.
\end{equation}
For later use we infer from equations (\ref{eq:(2.8)}\textit{-d}) the
useful relation
\begin{equation} \label{eq:useful}
    \begin{pmatrix}
        u_{0}v_{0} \\
        w_{0}z_{0}
    \end{pmatrix} = \frac{2}{\Delta}
    \begin{pmatrix}
        \hspace{2mm}    T_{22} & \hspace{-1mm} - T_{12} \\
        \hspace{-1mm} - T_{21} & \hspace{2mm}    T_{11}
    \end{pmatrix}
    \begin{pmatrix}
        \lambda \\
        \vartheta
    \end{pmatrix}
\end{equation}
with $\Delta = T_{11}T_{22}-T_{12}T_{21}$.  If accidentally
$\Delta=0$ happens then we may use the relation
$u_0v_0/w_0z_0=\mathit{const}$\, instead of \eref{eq:useful},
which case needs a special consideration.

\subsection{Recursion relations}

The next step of the P-analysis is to again substitute expansions
(\ref{eq:uvwz_sorf}\textit{-b}) and (\ref{eq:V_sorf}) with the
leading orders $p=q=r=s=-1$ into equations
(\ref{eq:GP_komp_a}\textit{-d}). After some algebra we obtain the
recursion relations
\numparts
\begin{equation} \label{eq:rekurzios_osszefugges}
    \hspace*{-1.5cm}
    \underbrace{\begin{pmatrix}
         Q_{1} & -T_{11}u_{0}^{2} & -T_{12}u_{0}z_{0} & -T_{12}u_{0}w_{0} \\
        -T_{11}v_{0}^{2} & Q_{1} & -T_{12}v_{0}z_{0} & -T_{12}v_{0}w_{0}\\
        -T_{21}v_{0}w_{0}& -T_{21}u_{0}w_{0} & Q_{2} & -T_{22}w_{0}^{2}  \\
        -T_{21}v_{0}z_{0}& -T_{21}u_{0}z_{0} & -T_{22}z_{0}^{2} & Q_{2}
    \end{pmatrix}}_{\displaystyle{Q(j)}}
    \begin{pmatrix}
        u_{j} \\ v_{j} \\ w_{j} \\ z_{j}
    \end{pmatrix} =
        \begin{pmatrix}
        F_{j} \\ G_{j} \\ H_{j} \\ K_{j}
    \end{pmatrix}
\end{equation}
where $j=1,2,\dots$ and
\begin{eqnarray} \label{eq:Q1Q2F}
   \fl Q_{1} = \lambda (j-1)(j-2) - 2T_{11}u_{0}v_{0} - T_{12}w_{0}z_{0},\\
   \fl Q_{2} = \vartheta  (j-1)(j-2) - 2T_{22}w_{0}z_{0} - T_{21}u_{0}v_{0},\\
   \label{eq:F_definition}
   \fl F_{j} = -\rmi u_{j-2,t} - \rmi (j-2)u_{j-1}\phi_{t} +
               \sum_{m=1}^{j-1}{(T_{11}u_{m}u_{j-m}v_{0} +
               T_{12}u_{m}w_{0}z_{j-m})} \nonumber \\
   \fl         + \sum_{l=1}^{j-1}{\sum_{m=0}^{l}{(T_{11}u_{m}u_{l-m}v_{j-l}
               + T_{12}u_{m}w_{j-l}z_{l-m})}} +
               \sum_{l=0}^{j-2}{V_{1,l}u_{j-l-2}} + V_{10,j-3} \, .
\end{eqnarray}
\endnumparts
Here we use the notation that whenever an index is less than zero,
then the expression itself is zero (for example, $V_{10,j-3}
\equiv 0$ for $j \le 2$). Furthermore $G_{j}$ is obtained from
$F_{j}$ by interchanging $u_{l}$ and $v_{l}$ and letting $\rmi
\rightarrow -\rmi$, $V_{1,l} \rightarrow V_{1,l}^{\ast}$ and
$V_{10,l} \rightarrow V_{10,l}^{\ast}$. The expressions $H_{j}$
and $K_{j}$ can be obtained, respectively, from $F_{j}$ and
$G_{j}$ by interchanging $u_{l}$ and $w_{l}$, $v_{l} $ and
$z_{l}$, $T_{12}$ and $ T_{21}$, $T_{11}$ and $T_{22}$\,, and
letting $V_{1} \rightarrow V_{2}$, $V_{10} \rightarrow V_{20}$.

The expressions $F_{j}$, $G_{j}$, $H_{j}$, $K_{j}$ at a given $j$
depend only on the expansion coefficients $u_{l}$, $v_{l}$,
$w_{l}$, $z_{l}$ with $l<j$. Therefore the equation
(\ref{eq:rekurzios_osszefugges}) represents recursion relations
for the determination of the unknowns $u_{j}$, $v_{j}$, $w_{j}$,
$z_{j}$ from the knowledge of the prior calculated coefficient
functions $u_{l}$, $v_{l}$, $w_{l}$, $z_{l}$ with $l<j$.

\subsection{Resonances \label{subsec:Resonances}}

The above recursion relations \eref{eq:rekurzios_osszefugges}
determine the unknown expansion coefficients uniquely unless the
determinant of the matrix $Q(j)$ is zero. Those values of $j$ at
which the determinant det($Q(j)$) becomes zero are called
resonances. After some calculation one obtains
\begin{eqnarray} \label{eq:csatolt_GP_detQ_altalanosan}
    \fl \det{\!\bigl ( Q(j) \bigr )} =
    \lambda^2 \vartheta^{2} (j+1) \,j^{2} (j-3)^{2} (j-4)\!
    \left ( j^{2} - 3j + 4-2\frac{\vartheta T_{11} u_{0}v_{0} +
    \lambda T_{22} w_{0}z_{0}}{\lambda \vartheta} \right ) \nonumber \\
\end{eqnarray}
so that the resonances of the coupled GP equations
(\ref{eq:GP_mat}\textit{-b}) are as follows
\begin{equation}
    j_{\mbox{\footnotesize{res}}} =
    \bigl \lbrace -1,0,0,3,3,4,j_{1},j_{2} \bigr \rbrace.
\end{equation}
Here $j_{1}$ and $j_{2}$ are the roots of the expression contained
in the last parentheses of equation
(\ref{eq:csatolt_GP_detQ_altalanosan}) and can formally be given
as
\begin{equation}\label{eq:j_12}
    j_{1,2} = \frac{3}{2} \pm \frac{1}{2}
              \sqrt{8\,\frac{\vartheta T_{11}u_{0}v_{0} +
              \lambda T_{22} w_{0}z_{0}}{\lambda \vartheta}-7
              \hspace*{1mm}} \quad \in \mathbb{Z}.
\end{equation}

As indicated, the resonances $j_{1}$ and $j_{2}$ must be integers
so that the square-root should be odd integers. From this one gets
a condition
\begin{equation} \label{eq:komp_par}
    \hspace*{-5mm}
    \sqrt{8\,\frac{\vartheta T_{11}u_{0}v_{0} +
    \lambda T_{22} w_{0}z_{0}}{\lambda \vartheta}-7
    \hspace*{1mm}} = 2m+1 \, , \hspace*{1cm} m = 0,1,2,
    \ldots
\end{equation}
involving the characteristic parameters $T_{ij}$, $\lambda$,
$\vartheta$ of the GP equations (\ref{eq:GP_mat}\textit{-b}). The
number $m$ can be considered as a classification number which
classifies possible external potential families for which the
system (\ref{eq:GP_mat}\textit{-b}) is integrable (in the general
sense of integrability \cite{Steeb1988,Ablowitz1991} ).

By re-arranging (\ref{eq:komp_par}) and using relation
(\ref{eq:useful}) one gets a more explicit condition necessary for
any coupled GP equations to pass the P-test:
\vspace*{2mm}
\begin{equation} \label{eq:integralhatosagi_foltetel_1}
   \fl
   \frac{2\,T_{11}T_{22}- \left (\vartheta/\lambda \right )\,
    T_{11} T_{12} - \left ( \lambda/\vartheta \right ) \, T_{21}
    T_{22}}{T_{11} T_{22}-T_{12}T_{21}} \!= \!\frac{1}{16}
    \bigl \lbrack (2m+1)^{2} + 7 \bigr \rbrack, \quad
    m = 0,1,2,\ldots
\end{equation}
\vspace*{2mm}
It is also clearly seen that this expression depends
only on the ratios $\lambda/\vartheta$, $T_{11}/T_{21}$, and
$T_{12}/T_{22}$ involving the characteristic parameters of the GP
equations.

In summary, any coupled system of GP equations
(\ref{eq:GP_mat}\textit{-b}) passes the Painlev{\'{e}} test only
if its characteristic parameters $\lambda$, $\vartheta$, $T_{ij}$
($i,j=1,2$) obey the relation
(\ref{eq:integralhatosagi_foltetel_1}), otherwise it is probably
not integrable. (See discussions about connection of P-test with
integrability in references \cite{Steeb1988,Ablowitz1991}.)

\subsection{Compatibility conditions
\label{subsec:kompatibilitasi_foltetelek}}

At each element of $j_{\mbox{\footnotesize{res}}}$, the recursion
relations (\ref{eq:rekurzios_osszefugges}) cannot be used for the
calculation of the expansion coefficients.  At these indices
arbitrary functions may arise in the expansions
(\ref{eq:uvwz_sorf}\textit{-b}). However, in order that the
solution be expressible in the form of the expansions
(\ref{eq:uvwz_sorf}\textit{-b}) and (\ref{eq:V_sorf}), the
recursion relations should be identically satisfied at $j \in
j_{\mbox{\footnotesize{res}}}$. The investigation of these
specific requirements leads to relations called compatibility
conditions which impose restrictions for the external potentials
$V_{i}(x,t)$ $i=1,2$. We note that only the positive resonances
are of interest.

\subsubsection{Compatibility condition belonging to resonance $j=3$.
\label{subsubseq:comp_cond_3}}

Let us consider equations (\ref{eq:rekurzios_osszefugges}) at $j=3$
and use equation (\ref{eq:useful}). The result is an equation
\begin{equation} \label{eq:j3_kompfeltetel_matrixos_alak}
    \hspace*{-1.3cm}
    \underbrace{\begin{pmatrix}
                    -T_{11}u_{0}v_{0} & -T_{11}u_{0}^{2}  &
                    -T_{12}u_{0}z_{0} & -T_{12}u_{0}w_{0}   \\
                    -T_{11}v_{0}^{2}  & -T_{11}u_{0}v_{0} &
                    -T_{12}v_{0}z_{0} & -T_{12}v_{0}w_{0}   \\
                    -T_{21}v_{0}w_{0} & -T_{21}u_{0}w_{0} &
                    -T_{22}w_{0}z_{0} & -T_{22}w_{0}^{2}    \\
                    -T_{21}v_{0}z_{0} & -T_{21}u_{0}z_{0} &
                    -T_{22}z_{0}^{2}  & -T_{22}w_{0}z_{0}   \\
                \end{pmatrix}}_{\displaystyle{Q(3)}}
                \begin{pmatrix}
                    u_{3} \\ v_{3} \\ w_{3} \\ z_{3}
                \end{pmatrix} =
                \begin{pmatrix}
                    F_{3} \\ G_{3} \\ H_{3} \\ K_{3}
                \end{pmatrix}
            \end{equation}
whose matrix $Q(3)$ has rank two. Indeed, by multiplying the first
row with $v_{0}$, the second with $u_{0}$, one gets a matrix which
possesses identical elements in its first two rows. Performing
similar manipulations, one can make the third and fourth rows also
identical. It then follows that the above recursion relation can
only be consistent if the following compatibility conditions hold
\numparts
\begin{eqnarray} \label{eq:j3_kompatibilitasi_foltetel}
                F_{3} v_{0} &=& G_{3} u_{0}, \\
                H_{3} z_{0} &=& K_{3} w_{0} \, .
\end{eqnarray}
\endnumparts
We emphasize that the above conditions are not independent from
each other because, for example as shown by \eref{eq:F_definition},
$F_{3}$ contains elements $w_{i}$, $z_{i}$ with $i \le 3$. Similarly,
$G_{3}$, $H_{3}$ and $K_{3}$ also contain all types of expansion
coefficients $u_{i}$, $v_{i}$, $w_{i}$, $z_i$ with $i \le 3$.

\subsubsection{Compatibility condition belonging to resonance $j=4$.}

By taking the recursion relations (\ref{eq:rekurzios_osszefugges})
at $j=4$ and applying equation (\ref{eq:useful}), one arrives at
the following equation
\begin{equation} \label{eq:j4_kompfeltetel_matrixos_alak}
   \fl \underbrace{\begin{pmatrix}
                   4\lambda -T_{11}u_{0}v_{0} & -T_{11}u_{0}^{2}  &
                   -T_{12}u_{0}z_{0} & -T_{12}u_{0}w_{0} \\
                   -T_{11}v_{0}^{2}  & 4\lambda -T_{11}u_{0}v_{0} &
                   -T_{12}v_{0}z_{0} & -T_{12}v_{0}w_{0} \\
                   -T_{21}v_{0}w_{0} & -T_{21}u_{0}w_{0} &
                   4\vartheta -T_{22}w_{0}z_{0} & -T_{22}w_{0}^{2}  \\
                   -T_{21}v_{0}z_{0} & -T_{21}u_{0}z_{0} &
                   -T_{22}z_{0}^{2}  & 4\vartheta-T_{22}w_{0}z_{0} \\
                \end{pmatrix}}_{\displaystyle{Q(4)}} \!\!
                \begin{pmatrix}
                    u_{4} \\ v_{4} \\ w_{4} \\ z_{4}
                \end{pmatrix} \!\!=\!\!
                \begin{pmatrix}
                    F_{4} \\ G_{4} \\ H_{4} \\ K_{4}
                \end{pmatrix}\!.
\end{equation}
In the general case the matrix $Q(4)$ has rank three which means
that only three of its rows are independent.  Using this fact,
after some calculation we obtain the following compatibility
condition
\begin{equation} \label{eq:j4_kompatibilitasi_foltetel1}
    T_{21} \Bigl ( F_{4}v_{0} + G_{4}u_{0} \Bigr ) +
    T_{12} \Bigl ( H_{4}z_{0} + K_{4}w_{0} \Bigr ) = 0.
\end{equation}

We should investigate also the possibility when $\mathrm{rank}
\bigl ( Q(4) \bigr) = 2$. In this case the compatibility
condition decomposes into two distinct parts as it can be seen
in the following way. The rank of a matrix equals the maximal
order of its non-singular submatrices. We should thus calculate
the determinants of all third order submatrices of $Q(4)$ and
investigate the cases when they simultaneously become zero. After
a simple but lengthy calculation the following results are
obtained for the determinants of the four third-order submatrices
of the matrix $Q(4)$:
\begin{equation}
   \hspace*{-0.8cm}
   \Bigl \lbrace 16 \lambda \vartheta T_{21}u_{0}z_{0}\,;\, -16
   \lambda \vartheta T_{12}u_{0}w_{0}\,;\, 16 \lambda \vartheta
   T_{12}w_{0}z_{0}\,;\, -16 \lambda \vartheta T_{21}u_{0}^{2}
   \Bigr \rbrace \, .
\end{equation}
Because $\lambda$ and $\vartheta$ are the non-zero mass parameters
[see definitions \eref{eq:parameters}], it is clear that the
subdeterminants vanish only if $T_{12}=T_{21}=0$.  This situation
however corresponds to two uncoupled GP equations.  We thus
arrived at the compatibility condition found by Clarkson
\cite{Clarkson1988} for the single GP equations: \numparts
\begin{eqnarray} \label{eq:special_4}
    F_{4} v_{0} + G_{4} u_{0} &=& 0, \\
    H_{4} z_{0} + K_{4} w_{0} &=& 0 \, .
\end{eqnarray}
\endnumparts

We note however that the general compatibility condition, as given
by equation (\ref{eq:j4_kompatibilitasi_foltetel1}), leads to more
complicated external potentials (discussed in the next section)
than that obtainable from equations
(\ref{eq:special_4}\textit{-b}) with $T_{12}=T_{21}=0$.

\subsubsection{Compatibility condition belonging to resonances $j_{1}$ and $j_{2}$.}

We now consider the matrix $Q(j_{1,2})$ with $j_{1,2}$ taken from
equation \eref{eq:j_12}. In general the matrix $Q(j_{1,2})$ has
rank three from which the following compatibility condition arises
\begin{equation} \label{eq:general_comp_j12}
    \vartheta w_{0} z_{0} \Bigl ( F_{j_{1,2}}v_{0} +
    G_{j_{1,2}}u_{0} \Bigr ) - \lambda u_{0} v_{0}
    \Bigl ( H_{j_{1,2}}z_{0} + K_{j_{1,2}}w_{0}
    \Bigr ) = 0 \, .
\end{equation}

As before we analyze also the case when $\mathrm{rank} \bigl (
Q(j_{1,2}) \bigr ) = 2$. After a lengthy but simple calculation we
obtain the determinants of the four third-order submatrices of the
matrix $Q(j_{12})$ to be:
\begin{eqnarray} \label{eq:submatrix_determinants_j12}
    \fl \Bigl \lbrace
        T_{21} \Delta^{5} \left ( u_{0}v_{0} \right )^{3}
        \left ( w_{0}z_{0} \right )^{2} z_{0} \,;\,
       -T_{21} \Delta^{5} \left ( u_{0}v_{0} \right )^{3}
        \left ( w_{0}z_{0} \right )^{2}         \,;
        \nonumber \hspace*{2cm}\\
        \hspace*{1.7cm}
        T_{21} \Delta^{5} \left ( u_{0}v_{0} \right )^{3}
        \left ( w_{0}z_{0} \right )^{2} z_{0} \,;\,
       -T_{12} \Delta^{6} \left ( u_{0}v_{0}w_{0}z_{0} \right )^{3}
       \Bigr \rbrace.
\end{eqnarray}

Now, as clearly seen the determinants
\eref{eq:submatrix_determinants_j12} vanish simultaneously only if
$T_{12}=T_{21}=0$ (or $\Delta=0$ which case is not considered
here). On the other hand, for this decoupled case one can
determine the values $j_{1}$ and $j_{2}$ by using definition
(\ref{eq:j_12}) and relation (\ref{eq:useful}) to be $j_{1}=4$,
$j_{2}=-1$. But then, as it can be checked easily by using
\eref{eq:rekurzios_osszefugges}, the corresponding compatibility
conditions reduce to those already discussed in connection with
equations (\ref{eq:special_4}-{\textit{b}}). We note however that,
depending on the experimental situations, it is possible to obtain
resonance values $j_{1}$ and $j_{2}$ greater than four. We should
then use equation \eref{eq:general_comp_j12} for drawing
conclusions about the admissible form of the external potentials.

\section{Possible forms of the external potentials}

In the preceding section we have found equations, called
compatibility conditions, that must be fulfilled in order that the
GP equations pass the P-test. In this section we exploit the
consequences of these equations for the general form of the
external potentials $V_{1}$, $V_{10}$, $V_{2}$, $V_{20}$ appearing
in equations (\ref{eq:csatolt_GP_fizikai_jelolessel}\textit{-b})
and (\ref{eq:GP_mat}\textit{-b}). The experimental realization of
such potentials may lead to detection of stable structures (like
vortices) in BEC.

Although the compatibility conditions are related with indices $j$
at which the recursion relations \eref{eq:rekurzios_osszefugges}
do not apply to the calculation of the unknown coefficients
$u_{j}(t)$, $v_{j}(t)$, $w_{j}(t)$, $z_{j}(t)$, the equations
(\ref{eq:j3_kompatibilitasi_foltetel}\textit{-b}),
(\ref{eq:j4_kompatibilitasi_foltetel1}) and
(\ref{eq:general_comp_j12}) can be reduced to expressions in which
only the zeroth order coefficient functions $u_{0}(t)$,
$v_{0}(t)$, $w_{0}(t)$, $z_{0}(t)$ are present.  That is because,
at $j \not \in j_{\mbox{\footnotesize{res}}}$, use of the
recursion relations \eref{eq:rekurzios_osszefugges} and the
definition \eref{eq:F_definition} of the functions $F_{j}$,
$G_{j}$, $H_{j}$, $K_{j}$, leads always to expansion coefficients
$u_{j}$, $v_{j}$, $w_{j}$, $z_{j}$ that are expressed by the
zeroth order functions $u_{0}(t)$, $v_{0}(t)$, $w_{0}(t)$,
$z_{0}(t)$.

In the following we shall present the results of the calculation
belonging to each compatibility condition.  For the resonance
$j=3$ the calculation can be performed easily by hand, but for
$j=4$ the computer program Maple \cite{Maple} had to be invoked in
order to perform the analytic manipulations. As a result of the
Maple program, all the coefficients which multiply the higher
order powers of $\phi_{t}$ proved to be analytically zero. The
expression associated with the zeroth order power of $\phi_{t}$
has been evaluated further by hand to get the final results which
will be presented and discussed below.

\subsection{Conditions for the potentials arising from $j=3$}

In section \ref{sec:Painleve_test} it has been shown that at $j=3$
the compatibility condition decomposes into two distinct parts
which are however not independent of each other [see equations
(\ref{eq:j3_kompatibilitasi_foltetel}\textit{-b}) and the remark
thereafter].

The elaboration of the compatibility conditions
(\ref{eq:j3_kompatibilitasi_foltetel}\textit{-b}) related with $j=3$
yields the relations
\numparts
\begin{eqnarray} \label{eq:comp_cond_expansion_3}
    \hspace*{-1.3cm}
    F_{3}v_{0} - G_{3}u_{0} =0 \quad &\longrightarrow & \quad
    (V_{1,1}-V_{1,1}^{\ast})u_{0}v_{0} + V_{10,0}v_{0} -
    V_{10,0}^{\ast}u_{0} = 0 \, , \\
    \hspace*{-1.3cm}
    H_{3}z_{0} - K_{3}w_{0} =0 \quad &\longrightarrow & \quad
    (V_{2,1}-V_{2,1}^{\ast})w_{0}z_{0} + V_{20,0}z_{0} -
    V_{20,0}^{\ast}w_{0} = 0 \, .
\end{eqnarray}
\endnumparts
It is clear from equation \eref{eq:useful} that only the products
$u_{0}v_{0}$ and $w_{0}z_{0}$ are determined uniquely so that one
element of each pair can be chosen arbitrarily. With the choices
$u_{0}=1$, $w_{0}=1$ the above relations can only be satisfied if
\numparts
\begin{eqnarray} \label{eq:j3_feltetel}
    V_{1,1}-V_{1,1}^{\ast} = 0 \qquad \mathrm{and} \qquad
    V_{10,0} = V_{10,0}^{\ast} \equiv 0, \\
    V_{2,1}-V_{2,1}^{\ast} = 0 \qquad \mathrm{and} \qquad
    V_{20,0} = V_{20,0}^{\ast} \equiv 0 .
\end{eqnarray}
\endnumparts
These conditions show that the expansion coefficients $V_{1,1}$ and
$V_{2,1}$ are real. Moreover, using  the definition \eref{eq:V_sorf}
for the expansion coefficients $V_{i,j}$ and the results
(\ref{eq:j3_feltetel}), we get
\begin{equation}
    0 = V_{10,0} = \frac{1}{0!} \left.
    \frac{\partial^{0} V_{10}(x,t)}{\partial x^{0}}
    \right \vert_{x= \xi(t)} \!\!= V_{10}( \xi(t),t).
\end{equation}
Since this equality holds for any arbitrary function $\xi(t)$, it
follows that $V_{10}(x,t)$ must vanish. Similar argumentation leads
to disappearance of $V_{20}(x,t)$.

In summary we conclude that in order that the equations
(\ref{eq:GP_mat}\textit{-b}) pass the P-test the inhomogeneity
terms must vanish and the first order expansion coefficient of the
external potentials should be real:
\numparts
\begin{eqnarray}
    V_{10} = V_{20} =0, \\
    V_{1,1} = V_{1,1}^{\ast} \quad \mbox{and}
    \quad V_{2,1} = V_{2,1}^{\ast} \, .
    \label{eq:V11_V21_reality}
\end{eqnarray}
\endnumparts

\subsection{Conditions for the potentials arising from $j=4$}

Without presenting the details of the algebraic manipulations, we
state that the compatibility condition
\eref{eq:j4_kompatibilitasi_foltetel1} (partly with the aid of the
formula manipulation program Maple) leads to the relation
\begin{eqnarray} \label{eq:j4-es_feltetel}
   \fl \frac{T_{21}}{2\lambda} u_{0} v_{0}
    (V_{1,0}-V_{1,0}^{\ast})^{2} +
    \frac{T_{12}}{2\vartheta} w_{0} z_{0}
    (V_{2,0}-V_{2,0}^{\ast})^{2}
    \nonumber \\
    - T_{12}w_{0}z_{0} (V_{2,2}+V_{2,2}^{\ast})
    - T_{21}u_{0}v_{0} (V_{1,2}+V_{1,2}^{\ast})
    \nonumber \\
    - \rmi \frac{T_{21}}{2\lambda} u_{0} v_{0}
    \parcdiff{}{t} (V_{1,0}-V_{1,0}^{\ast})
    - \rmi \frac{T_{12}}{2\vartheta} w_{0} z_{0}
    \parcdiff{}{t} (V_{2,0}-V_{2,0}^{\ast}) = 0
    \label{eq:j4_kompatibilitasi_foltetel2}
\end{eqnarray}
in which, as expected, only the zeroth order coefficient functions
$u_{0}(t)$, $v_{0}(t)$, $w_{0}(t)$, $z_{0}(t)$ appear together
with the parameters $\lambda$, $\vartheta$ and $T_{ij}$
$(i,j=1,2)$ in a special combination. This condition looks much
more complicated than that obtained above [cf. with equations
(\ref{eq:comp_cond_expansion_3}\textit{-b})]. Moreover both
potentials $V_{1}$ and $V_{2}$ are occurring within a single
relation.

Let us now write the external potentials in the form
\numparts
\begin{eqnarray} \label{eq:j4_potencial_valos_kepzetes_szetvalasztas}
     V_{1} (x,t) &=& \alpha (x,t) + \rmi \beta (x,t), \\
     V_{2} (x,t) &=& \gamma (x,t) + \rmi \delta (x,t)
\end{eqnarray}
\endnumparts
where $\alpha$, $\gamma$ and $\beta$, $\delta$ are real functions.
Exploiting the reality of $V_{1,1}$ and $V_{2,1}$ expressed by
relation \eref{eq:V11_V21_reality} and using the definition
\eref{eq:V_sorf}, we get the results
\begin{equation} \label{eq:beta_delta}
    \beta(x,t) \equiv \beta(t)
    \qquad \mathrm{and} \qquad
    \delta(x,t) \equiv \delta(t).
\end{equation}
This condition which can be checked easily by direct substitution
tells that the imaginary part of the potential may depend only on
the time $t$ but not on the space $x$ variables. Using this last
result and inserting the definitions
(\ref{eq:j4_potencial_valos_kepzetes_szetvalasztas}\textit{-b})
into the relation (\ref{eq:j4-es_feltetel}) we get the following
expression
\begin{eqnarray}
    \fl \left \lbrack
    - \frac{2}{\lambda}   T_{21} u_{0}v_{0} \beta^{2}
    - \frac{2}{\vartheta} T_{12} w_{0}z_{0} \delta^{2}
    + \frac{1}{\lambda}   T_{21} u_{0}v_{0} \totdiff{\beta}{t}
    + \frac{1}{\vartheta} T_{12} w_{0}z_{0} \totdiff{\delta}{t}
    \right \rbrack \nonumber \\
    \hspace*{4.3cm}
    - T_{12} w_{0}z_{0} \sparcdiff{\gamma}{x}
    - T_{21} u_{0}v_{0} \sparcdiff{\alpha}{x} = 0
    \label{eq:j4_kompatibilitasi_foltetel3} \, .
\end{eqnarray}
Because  the quantity in the square bracket depends only on time $t$,
integration by $x$ twice yields the following results
\begin{equation} \label{foeredmeny1}
    T_{12} w_{0}z_{0} \gamma + T_{21} u_{0}v_{0} \alpha =
    C_{1}(t) x^{2} + C_{2}(t) x + C_{3}(t)
\end{equation}
where the coefficients $C_{1}(t)$, $C_{2}(t)$, $C_{3}(t)$ depend
only on time $t$, and $C_{2}$, $C_{3}$ are arbitrary real
functions. By re-substituting this latter equation into expression
(\ref{eq:j4_kompatibilitasi_foltetel3}), we get the  constraint
for the function $C_{1}(t)$ as follows
\begin{equation} \label{foeredmeny2}
    C_{1} (t) = \frac{T_{21}}{\lambda} u_{0}v_{0} \left (
                \frac{1}{2} \totdiff{\beta}{t} - \beta^{2}
                \right ) +
                \frac{T_{12}}{\vartheta} w_{0}z_{0} \left (
                \frac{1}{2} \totdiff{\delta}{t} - \delta^{2}
                \right ) \, .
\end{equation}

We emphasize that the above result does not mean a restriction for
the individual form of the real part of the external potentials
$V_{1}$ and $V_{2}$. As equation (\ref{foeredmeny1}) shows only a
weighted sum of the real parts $\alpha$ and $\gamma$ is
constrained by the compatibility conditions
(\ref{eq:j4-es_feltetel}) belonging to the resonance $j=4$.

Let us now exhibit a possible consequence of the general constraints
(\ref{foeredmeny1}) and (\ref{foeredmeny2}) for the potentials by starting
from an obvious splitting of the coefficient $C_1(t)$ as follows
\begin{equation}
    C_{1}(t) = C_{1}^{(1)}(t) + C_{1}^{(2)}(t)
\end{equation}
with
\numparts
\begin{equation}
    \fl C_{1}^{(1)} = \frac{T_{21}u_{0}v_{0}}{\lambda} \left (
                      \frac{1}{2} \totdiff{\beta}{t} - \beta^{2}
                      \right )
        \quad \mathrm{and} \quad
        C_{1}^{(2)} = \frac{T_{12}w_{0}z_{0}}{\vartheta} \left (
                      \frac{1}{2} \totdiff{\delta}{t} - \delta^{2}
                      \right ).
\end{equation}
\endnumparts
Then the compatibility condition expressed by equation
(\ref{foeredmeny1}) can be satisfied by the following choices:
\numparts
\begin{eqnarray}
    T_{12} w_{0} z_{0} \gamma &=&
    C_{1}^{(1)} x^{2} + C_{2}^{(1)} x + C_{3}^{(1)} + f(x,t) \\
    T_{21} u_{0} v_{0} \alpha &=&
    C_{1}^{(2)} x^{2} + C_{2}^{(2)} x + C_{3}^{(2)} - f(x,t)
\end{eqnarray}
\endnumparts
where $C_{i}^{(1)}$, $C_{i}^{(2)}$ $(i=2,3)$ are arbitrary real
functions of time $t$ and $f$ is an arbitrary real function of
$x$ and $t$. Using the expressions
(\ref{eq:j4_potencial_valos_kepzetes_szetvalasztas}\textit{-b}) we get the
following possible form for the external potentials:
\numparts
\begin{eqnarray} \label{eq:potencial_alakok}
    \fl V_{1} \!&=&\! \frac{1}{\lambda} \!\left ( \frac{1}{2}
                  \totdiff{\beta(t)}{t} - \beta^{2}(t)\!
                  \right ) \!x^{2} + V_{1}^{(1)}(t) x +
                  V_{1}^{(0)}(t) - \frac{\tilde{V}(x,t)}{T_{21}
                  \left ( \lambda T_{22} - \vartheta T_{12}
                  \right )} + \rmi \beta(t) \, ,\\
    \fl V_{2} \!&=&\! \frac{1}{\vartheta} \!\left ( \frac{1}{2}
                  \totdiff{\delta(t)}{t} - \delta^{2}(t)
                  \right ) \!x^{2} + V_{2}^{(1)}(t) x +
                  V_{2}^{(0)}(t) + \frac{\tilde{V}(x,t)}{T_{12}
                  \left ( \vartheta T_{11} - \lambda T_{21}
                  \right )} + \rmi \delta(t)
\end{eqnarray}
\endnumparts
where $V_{1}^{(1)}$, $V_{1}^{(0)}$, $V_{2}^{(1)}$ and $V_{2}^{(0)}$
are arbitrary real functions of time $t$ and $\tilde{V}(x,t)$
represents an arbitrary real potential function which may be used
conveniently in design of experiments with BEC. At this point we
have to note that these formulae cannot be used for the uncoupled
case, since when $T_{12}=T_{21}=0$, then the compatibility condition
\eref{eq:j4_kompatibilitasi_foltetel1} changes to
(\ref{eq:special_4}\textit{-b}),
and in this way $\tilde{V}(x,t)$ does not arise.

In summary we conclude that in order that the coupled GP equations
\mbox{(\ref{eq:GP_mat}-\textit{b})} pass the P-test, a special
combination of the real parts of the potentials $V_{1}$ and
$V_{2}$ may depend only quadratically and/or linearly on the
spatial coordinate $x$. A stringent relationship can be
established between the coefficient of the quadratic terms and the
imaginary parts which, in turn, may depend only on time $t$. An
additional real potential $\tilde{V}$ of general form may be
introduced which explicitly exhibits coupling between the external
potentials $V_{1}$ and $V_{2}$.

\section{Discussion of the results}

In this section we discuss the results from various points of view
and make comparison with related results obtained by others.

\subsection{Presence of source terms}

In the course of the theoretical study of two-component BEC with
attractive in\-ter\-ac\-tion, it has been found
\cite{Adhikari2000} that the decay and growth of number of atoms
is best accounted for by introducing an imaginary contact
interaction term in the GP equations. We now see that our analysis
enables the existence of such source terms, by appropriately
chosen $\beta(t)$ and $\delta(t)$ [see equation
(\ref{eq:potencial_alakok}-{\textit{b})]. This result holds also
in  the case of one-component BEC as found by Clarkson
\cite{Clarkson1988}.

\subsection{Uncoupled case}

Next, we investigate the case $T_{12}=T_{21} \equiv 0$, when the
system of the GP equations (\ref{eq:GP_mat}\textit{-b}) is
decoupled. As an example we derive the resonances. Our general
equations should reduce twice to earlier results obtained by
Clarkson \cite{Clarkson1988} for the one-component GP equation.
Starting from the general expression
(\ref{eq:csatolt_GP_detQ_altalanosan}) and applying the useful
formula (\ref{eq:useful}) one obtains
\begin{eqnarray}
    \fl \det{ \bigl ( Q (j) \bigr )} =
                    \lambda^2 \vartheta^{2} (j+1) \,j^{2} (j-3)^{2}
                    (j-4) \left ( j^{2}-3j+4-2 \frac{T_{11}T_{22}}{\Delta}
                    \frac{2 \lambda \vartheta + 2 \lambda \vartheta}{\lambda
                    \vartheta} \right ) \nonumber \\
    \fl             = \lambda^2 \vartheta^{2} (j+1) \,j^{2} (j-3)^{2}
                    (j-4) \left ( j^{2} - 3j - 4 \right ) =
                    \lambda^{2} \vartheta^{2} (j+1)^{2} \,j^{2}
                    (j-3)^{2} (j-4)^{2}\, .
\end{eqnarray}
The resonances $(-1,0,3,4)$ are those found by Clarkson \cite{Clarkson1988}
and all have a multiplicity of two as a result of the double number of the
(uncoupled) GP equations.

\subsection{Sign of the potential}

One of the experimental situations where the coupled GP equations
(\ref{eq:GP_mat}\textit{-b}) serve as a theoretical basis is the
creation of two component BEC \cite{Matthews1999}. In such
experiments alkali atoms are confined by symmetrically arranged
harmonic trap potentials. One of the possibility of our results is
to choose in equations (\ref{eq:potencial_alakok}\textit{-b}) all
functions $V_{1}^{(1)}$, $V_{1}^{(0)}$, $V_{2}^{(1)}$,
$V_{2}^{(0)}$, $\beta$, $\delta$ equal to zero and let the
potential $\tilde{V}(x,t)$ operate as a field confining the alkali
gas particles. It is then required that in equations
(\ref{eq:potencial_alakok}\textit{-b}) the terms in which our
confining potential $\tilde{V}$ occurs do have the same sign. The
condition that those two terms with $\tilde{V}$ have the same sign
is in general
\numparts
\begin{equation}
    T_{12} T_{21} \left ( \lambda T_{22} - \vartheta T_{12} \right )
    \left ( \vartheta T_{11} - \lambda T_{21} \right ) < 0
\end{equation}
which can be expressed also in terms of the scattering lengths as
\begin{equation}
    a_{12} a_{21} \left ( \lambda a_{22} - \vartheta a_{12} \right )
    \left ( \vartheta a_{11} - \lambda a_{21} \right ) < 0 \, .
\end{equation}
\endnumparts
Because, physically $a_{12}=a_{21}$, the above condition for the
equality of signs of the $\tilde{V}$ terms in equations
(\ref{eq:potencial_alakok}-{\textit{b}}) is fulfilled for the
usual experimental case with $ \lambda=\vartheta $ if
\begin{equation} \label{eq:sign_condition}
   a_{11} < a_{12} < a_{22}
   \qquad \mathrm{or} \qquad
   a_{22} < a_{12} < a_{11}.
\end{equation}
If the scattering lengths $a_{12}=a_{21}$ are greater or lesser
than both $a_{11}$ and $a_{22}$ then the sign of the terms
containing the arbitrary potential $\tilde{V}$ is different, which
corresponds to untrapping one of the BEC components.

\subsection{Fulfillment of equation \eref{eq:integralhatosagi_foltetel_1}}

The best studied example of the two-component BEC involves Rb
atoms in two different hyperfine states. It has been found
experimentally \cite{Hall1998_1,Hall1998_2,Matthews1999}, and
numerically \cite{Garcia1999} that a stable configuration of
soliton-like vortex in the two-component condensate is achieved in
the case where the scattering lengths are in the proportion:
\begin{equation} \label{eq:ratios}
    a_{11}\, : \,a_{12}\, : \,a_{22}\, = 1.03 :  1 :  0.97,
    \qquad \mbox{with} \quad
    a_{12} \equiv a_{21} \, .
\end{equation}
Let us now check whether these ratios obey our general condition
(\ref{eq:integralhatosagi_foltetel_1}) with integer $m$. Since
$\lambda/\vartheta=1$ expression
(\ref{eq:integralhatosagi_foltetel_1})
can be written as follows
\begin{equation}
   \fl
      \frac{2(a_{11}/a_{21}) (a_{22}/a_{12}) -
           (a_{11}/a_{21})-(a_{22}/a_{12})}{(a_{11}/a_{21})(a_{22}/a_{12})-1}
    = \frac{1}{16} \bigl \lbrack (2m+1)^{2}+7 \bigr \rbrack \, .
\end{equation}
The insertion of the above ratios gives
\begin{equation}
    \fl
    \frac{2\cdot 1.03\cdot 0.97 - 1.03 - 0.97}
    {1.03\cdot 0.97 - 1} = \frac{-0.0018}{-0.0009} = 2
    \equiv \frac{1}{16} \bigl \lbrack (2m+1)^{2} + 7 \bigr \rbrack
\end{equation}
which yields
\begin{equation}
    \hspace*{3.5cm} m = 2.
\end{equation}
This result means that the experimental ratios (\ref{eq:ratios})
correspond to just one of the possible solutions of the GP
equations characterized by a $m=2$ potential family.

Proceeding further, one can determine the resonances belonging to
the ex\-per\-i\-men\-tal\-ly found ratios (\ref{eq:ratios}) to be
[cf. with equations (\ref{eq:j_12}), (\ref{eq:komp_par})]
\begin{equation}
    j_{1}=2+m=4, \hspace*{3cm} j_{2} = 1-m = -1.
\end{equation}

This result means that no further work is needed, the underlying
potential falls into the category defined by the compatibility
condition for $j=4$; a possible representation of such potentials
is given by (\ref{eq:potencial_alakok}\textit{-b}). Indeed, the
quadratic trap potential used in the experiments suits well to the
general form of potentials obtained from the analysis of the
resonance at $j=4$.

\section{Summary}

In this paper the first step towards verification of integrability
of the coupled GP equations by means of the P-analysis has been
carried out. It has been shown that the GP equations pass the
P-test provided a special relation among the system parameters
(masses, interaction strengths) is satisfied [cf. with
\eref{eq:integralhatosagi_foltetel_1}]. One of the recent
experiments has been taken as an example. In this experiment
\cite{Hall1998_1,Hall1998_2,Matthews1999} and a subsequent
numerical study \cite{Garcia1999}, the vortex stability of a two
component BEC has been investigated. It is found that the system
parameters at which stability occurs are just in the proportion
which fits our relation \eref{eq:integralhatosagi_foltetel_1} with
$m=2$, a condition necessary for the GP equation to pass the
P-test.

As the GP-equations play a great role in describing BEC, a
particular attention has been paid to establish the admissible
forms of the confining trap potentials of experimental interest.
It has been found that, in addition to the prescribed form
resulted by the P-analysis of a single GP equation, there is a
possibility of introducing an extra potential term of arbitrary
shape into the external potentials [cf. with equations
(\ref{eq:potencial_alakok}-{\textit{b}})]. Also, some discussion
of the results has been added which includes the comparison of the
earlier results obtained for the one component GP equation, the
role of the source (imaginary) terms $\beta(t)$ and $\delta(t)$ in
the potentials, and the sign of the additional potential terms.

Finally, we add a remark to the fulfillment of equation
\eref{eq:integralhatosagi_foltetel_1} with integer $m$'s. In the
light of experimental errors the above agreement $m=2$ may seem to
be accidental. We note however that soliton-type structures (e.g.
vortices in 3D) possess an outstanding stability sometimes called
'robustness'  which enables these particle-like formations to
survive for a long time or even to arise in circumstances that do
not fit the exact constraint of mathematics. Therefore equation
\eref{eq:integralhatosagi_foltetel_1} may prove also useful in
exploring other regions of parameters where such stable structures
are to be observed in binary condensates.

\ack We would like to thank professor Istv{\'{a}}n Nagy for
fruitful discussions about excitations in BEC. This research was
supported by OTKA under Grant Nos. T29884, T25019.

%
%


\section*{References}

\begin{thebibliography}{99}

    \bibitem{Gross1961}
        Gross E P 1961 \NC {\textbf{20}} 454

    \bibitem{Pitaevskii1961}
        Pitaevskii L P 1961 {\textit{Soviet Phys -- JETP}} {\textbf{40}} 646

%
%
    \bibitem{Ho1996}
        Ho Tin-Lun and Shenoy V B 1996 \PRL {\textbf{77}} 3276--9

    \bibitem{Ballagh1997}
        Ballagh R J, Burnett K and Scott T F 1997 \PRL {\textbf{78}} 1607--11

    \bibitem{Esry1997}
      Esry B D, Greene C H, Burke J P Jr. and Bohn J L 1997 \PRL {\textbf{78}} 3594--7

    \bibitem{Hisakado1997}
      Hisakado M 1997 {\textit{Preprint}} solv-int/9701022v2

    \bibitem{Hall1998_1}
        Hall D S, Matthews M R, Ensher J R, Wieman C E and Cornell E A 1998 \PRL {\textbf{81}} 1539--42

    \bibitem{Hall1998_2}
        Hall D S, Matthews M R, Wieman C E and Cornell E A 1998 \PRL {\textbf{81}} 1543--6

    \bibitem{Villain1998}
        Villain P and Lewenstein M 1998 {\textit{Preprint}} quant-ph/9808017

    \bibitem{Williams99}
        Williams J, Walser R, Cooper J, Cornell E and Holland M 1999 {\textit{Phys. Rev. A.}} {\textbf{59}} R31-4

    \bibitem{Raghavan99}
        Raghavan S, Smerzi A and Kenkre V M 1999 {\textit{Phys. Rev. A.}} {\textbf{60}} R1787-90

    \bibitem{Ho98}
        Ho T L 1998 \PRL {\textbf{81}} 742-5

    \bibitem{Law81}
        Law C K, Pu H and Bigelow N P 1998 \PRL {\textbf{81}} 5257-61

    \bibitem{Pu99}
        Pu H, Raghavan S and Bigelow N P 2000 {\textbf{61}} 023602

    \bibitem{Pu00}
        Pu H, Law C K, Raghavan S, Eberly J H and Bigelow N P 1999
        {\textit{Phys. Rev. A.}} {\textbf{60}} 1463-70

    \bibitem{Matthews1999}
        Mattews M R, Anderson B P, Haljan P C, Hall D S, Wieman D S and Cornell E A 1999 \PRL {\textbf{83}} 2498--501

    \bibitem{Narayanan1999}
      Narayanan A, Ramachandran H 1999 {\textit{Preprint}} cond-mat/9911390v2

    \bibitem{Garcia1999}
        P\`{e}rez-Garc{\'\i}a V M and Garc{\'\i}a-Ripoll J J 1999 {\textit{Preprint}} cond-mat/9912308

    \bibitem{Ohberg1999}
      {\"O}hberg P and Stenholm S 1999 \JPB {\textbf{32}} 1959--70

    \bibitem{Abdullaev2000_1}
       Abdullaev F Kh, Kraenkel R A 2000 {\textit{Preprint}} cond-mat/0004117

    \bibitem{Abdullaev2000_2}
       Abdullaev F Kh, Kraenkel R A 2000 {\textit{Preprint}} cond-mat/0005445

    \bibitem{Adhikari2000}
        Adhikari S K 2000 {\textit{Preprint}} cond-mat/0011526v2

%
%

    \bibitem{Hasegawa1990}
        Hasegawa A 1989 {\textit{Optical Solitons in Fibers}} (Springer-Verlag, Heidelberg)

    \bibitem{Ruprecht1995}
        Ruprecht P A, Holland M J and Burnett K 1995 {\textit{Phys. Rev. A}} {\textbf{51}} 4704--11

    \bibitem{Heinzen2000}
        Heinzen D J, Wynar R, Drummond P D and Kheruntsyan K V 2000 \PRL {\textbf{84}} 5029--33

   \bibitem{Weiss1983}
        Weiss J, Tabor M, Carnevale G 1983 \JMP {\textbf{24}} 522--6

    \bibitem{Steeb1988}
        Steeb W H and Euler N 1988 {\textit{Nonlinear evolution equations
        and Painlev{\'{e}} test}} (Singapore: World Scientific)

    \bibitem{Ablowitz1991}
        Ablowitz M J and Clarkson P A 1991 {\textit{Solitons,
        Nonlinear Evolution Equations and Inverse Scattering}}
        (Cambridge: Cambridge University Press)

    \bibitem{Steeb1984}
        Steeb W-H, Kloke M and Spieker B-M 1984 \JPA {\textbf{17}} L825--9

    \bibitem{Clarkson1988}
        Clarkson P A 1988 {\textit{Proc. Roy. Soc. Edin.\,A.}} {\textbf{109A}} 109--26

    \bibitem{Radhakrishnan1995}
        Radhakrishnan R, Lakshmanan M and Daniel M 1995 \JPA {\textbf{28}} 7299--314

    \bibitem{Sakovich2000}
        Sakovich S Yu, Tsuchida T 2000 \JPA {\textbf{33}} 7217--26

    \bibitem{Fortagh1998}
        Fort{\'{a}}gh J, Grossmann A, Zimmermann C and H{\"{a}}nsch T W 1998 \PRL {\textbf{81}} 5310--3

    \bibitem{Maple}
        Heck A 1996 {\textit{Introduction to Maple}} (Berlin: Springer Verlag)

\end{thebibliography}
\end{document}